\documentclass{PoS}

\usepackage{physics}
\usepackage{booktabs}

\DeclareMathAlphabet{\pazocal}{OMS}{zplm}{m}{n}
\newcommand{\dint}{\mathrm d}
\newcommand{\Deff}{D_{\mathrm{eff}}}

\title{Lattice QCD estimate of the quark-gluon plasma photon emission rate}

\ShortTitle{Lattice QCD estimate of the quark-gluon plasma photon emission rate}

\author{Bastian B. Brandt,$^a$ Marco C{\`e},$^d$\footnote{Current affiliation: CERN, Theoretical Physics Department, 1211 Geneva 23, Switzerland}  Anthony Francis,$^b$ Tim Harris,$^c$ Harvey B. Meyer,$^{d,e}$ Aman Steinberg,$^f$ \speaker{Arianna Toniato}$^e$ \\
	\llap{$^a$} Institute for Theoretical Physics, Goethe University, Max-von-Laue-Strasse 1, 60438 Frankfurt am Main, Germany\\
	\llap{$^b$} CERN, Theoretical Physics Department, 1211 Geneva 23, Switzerland\\
	\llap{$^c$} Dipartimento di Fisica, Universit{\`a} di Milano-Bicocca, and INFN, sezione di Milano-Bicocca, Piazza della Scienza 3, I-20126 Milano, Italy\\
	\llap{$^d$} Helmholtz-Institut Mainz, Johannes Gutenberg-Universit{\"a}t Mainz, D-55099 Mainz, Germany\\
	\llap{$^e$} PRISMA+ Cluster of Excellence \& Institut f{\"u}r Kernphysik, Johannes Gutenberg-Universit{\"a}t Mainz, D-55099 Mainz, Germany\\
	\llap{$^f$} Fakult{\"a}t f{\"u}r Physik, Universit{\"a}t Bielefeld, D-33615 Bielefeld, Germany\\
        E-mail: \email{toniato@uni-mainz.de}}

\abstract{We present a computation of the photon emission rate of the quark-gluon plasma from two-flavor lattice QCD at a temperature of 254 MeV, which follows up on the work presented in \cite{Brandt:2017vgl}. We perform a continuum extrapolation of the vector-current correlator, and consider a linear combination of the Lorentz indices corresponding to a UV-finite spectral function. To extract the spectral function from the lattice correlators, an ill-posed inverse problem, we model the spectral function with a Pad{\'e} ansatz. We further constrain our analysis by simultaneously fitting data with different momenta. We present results for a multi-momentum fit including the three smallest momenta available from our lattice analysis. \\ \flushright{\hfill MITP/19-082}}

\FullConference{37th International Symposium on Lattice Field Theory - Lattice2019\\
		16-22 June 2019\\
		Wuhan, China}

\begin{document}

\section{Introduction}

The quark-gluon plasma, the high-temperature phase of QCD matter, has been the object of extended experimental and theoretical interest over the past forty years. One of the observables carrying information on its properties is the photon emission rate. Thermal photons emitted by the quark-gluon plasma form part of the total photon spectrum measured in heavy-ion collisions. The size of the hot thermal medium generated in the collision is typically much smaller than the mean free path of photons, which undergo no further scattering after being produced and therefore carry direct information on the medium. A lattice computation of the photon emission rate in two-flavor QCD  at a temperature of 254 MeV from continuum-extrapolated data has been presented in \cite{Brandt:2017vgl}. In these proceedings, we present a follow-up to that work. In particular, in the present analysis we improve the quality of the continuum extrapolation by introducing a new fine lattice ensemble, and we make use of a more constraining fitting strategy for the computation of the final results. 


\section{Photon rate and spectral function}
\label{definitions}

In this section, we introduce the relevant observables, following closely reference \cite{Brandt:2017vgl}, to which we refer for a more detailed discussion. In Euclidean space, we consider the correlator
\begin{equation}
G_{\mu \nu}(x_0, \textbf k) = \int \dint^3 x \; e^{-i \: \textbf k \cdot \textbf x} \langle j_{\mu} (x) j_{\nu}^{\dagger} (0) \rangle \; ,
\label{corr}
\end{equation}
where $j_{\mu} (x) = \sum_f Q_f \bar \psi_f (x) \gamma_{\mu} \psi_f (x)$ is the electromagnetic current. The spectral representation of the correlator relates it to the spectral function $\rho_{\mu\nu}(\omega, \textbf k)$ via
\begin{equation}
G_{\mu \nu}(x_0, \textbf k) \underset{\mu\nu \neq 0i}{=} \int \frac{\dint \omega}{2 \pi} \; \rho_{\mu\nu}(\omega, \textbf k) \; \frac{\cosh \bigl[ \omega (\beta/2 - x_0) \bigr]}{\sinh (\omega \beta/2)}  \; ,
\label{spectral_rep}
\end{equation}
where $\beta \equiv 1/T$ is the inverse temperature.
The photon emission rate per unit volume of the quark-gluon plasma, at leading order in the electromagnetic coupling $e$, can be expressed in terms of $\rho_{\mu\nu}$ as \cite{McLerran:1984ay}
\begin{equation}
\dint \Gamma (\textbf k) = e^2 \frac{\dint^3 k}{(2\pi)^3 2\abs{\textbf k}} \frac{ \rho_{ii}(\omega = \abs{\textbf k}, \textbf k) - \rho_{00}(\omega = \abs{\textbf k}, \textbf k)}{e^{\beta \abs{\textbf k}} - 1} \; .
\label{phora}
\end{equation}
We consider the linear combination
\begin{equation}
\rho (\omega, \textbf k ; \lambda) \equiv \biggl( \delta_{ij} - \frac{k_i k_j}{\abs{\textbf k}^2} \biggr) \rho_{ij} (\omega, \textbf k) + \lambda \biggl(  \frac{k_i k_j}{\abs{\textbf k}^2} \: \rho_{ij} (\omega, \textbf k) - \rho_{00}(\omega, \textbf k) \biggr) \: ,
\label{rho_lambda}
\end{equation}
which for $\lambda = 1$ corresponds to the expression required to compute the photon rate, $\rho (\omega, \textbf k ; 1) = \rho_{ii} (\omega, \textbf k) - \rho_{00} (\omega, \textbf k)$. As a consequence of current conservation, for $\omega = \abs{\textbf k}$ the right-hand side of equation~(\ref{rho_lambda}) is independent of $\lambda$ \cite{Brandt:2017vgl},
implying that the term $(\rho_{ii} - \rho_{00}) \vert_{\omega = \abs{\textbf k}}$ in equation~(\ref{phora}) can be replaced by $\rho (\abs{\textbf k}, \textbf k ; \lambda)$ with arbitrary $\lambda$. In particular, the choice $\lambda =-2$ has very interesting properties: due to Lorentz invariance and transversality, $\rho (\omega, \textbf k ; \lambda = -2)$ vanishes identically in the vacuum, and it is UV-finite at finite temperature \cite{Brandt:2017vgl}. Moreover, the spectral function with $\lambda = -2$ is non-negative for $\omega \leq \abs{\textbf k}$, and a super-convergent sum rule can be derived for it \cite{Brandt:2017vgl}, which reads
\begin{equation}
\int_0^{\infty} \dint \omega \: \omega \: \rho (\omega, \textbf k; \lambda = -2) = 0 \: .
\label{sum_rule}
\end{equation}
%

\section{Lattice setup}

To compute the photon emission rate of the quark-gluon plasma in two-flavor lattice QCD, we make use of several lattice ensembles with a fixed temperature of $T = 254~\mathrm{MeV}$, generated with the Wilson gauge action and non-perturbatively $O(a)$-improved Wilson fermions. The running of the quark mass and gauge coupling, used to tune the bare parameters, was determined by the ALPHA collaboration \cite{Fritzsch:2012wq}. Details on the ensembles are reported in table \ref{ensembles}.

\begin{table}
     \center
     \begin{tabular}{cccccccc}
     \toprule
     label & $T$ (MeV) & $6/g_0^2$ & $\kappa$ & $\beta/a$ & $L/a$ & $m_{\overline{\mathrm{MS}}(2~\mathrm{GeV})}$ (MeV) & $N_{\mathrm{conf}}$ \\
     \midrule
     F7 & 254(5) & 5.3 & 0.13638 & 12 & 48 & 13 & 482 \\
     O7 & '' & 5.5 & 0.13671 & 16 & 64 & 13 & 305 \\
     W7 & '' & 5.68573 & 0.136684 & 20 & 80 & 13 & 1566 \\
     X7 & '' & 5.82716 &  0.136544 & 24 & 96 & 16 & 511 \\
     \bottomrule
     \end{tabular}
     \caption{Ensembles used in this work. The ensemble W7 is new and has been generated using openQCD version 1.6, while the others were part of the work presented in \cite{Brandt:2017vgl}. The coarsest ensemble F7 is excluded from the continuum extrapolation. The parameters of F7 and O7 match the ones of corresponding zero-temperature ensembles from the CLS initiative.}
     \label{ensembles}
\end{table}

\begin{figure}
\center
\includegraphics[width=.5\textwidth]{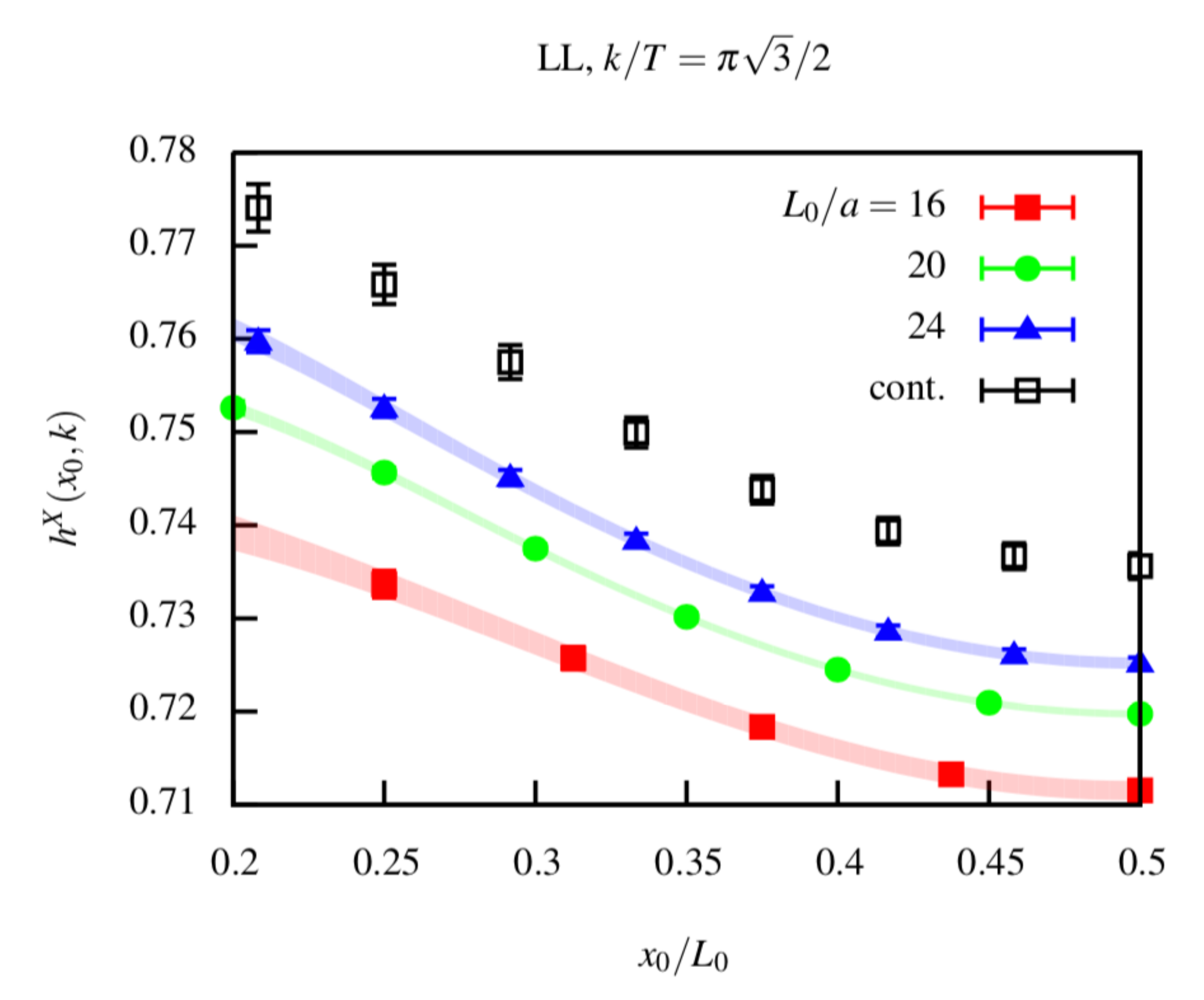}~\includegraphics[width=.5\textwidth]{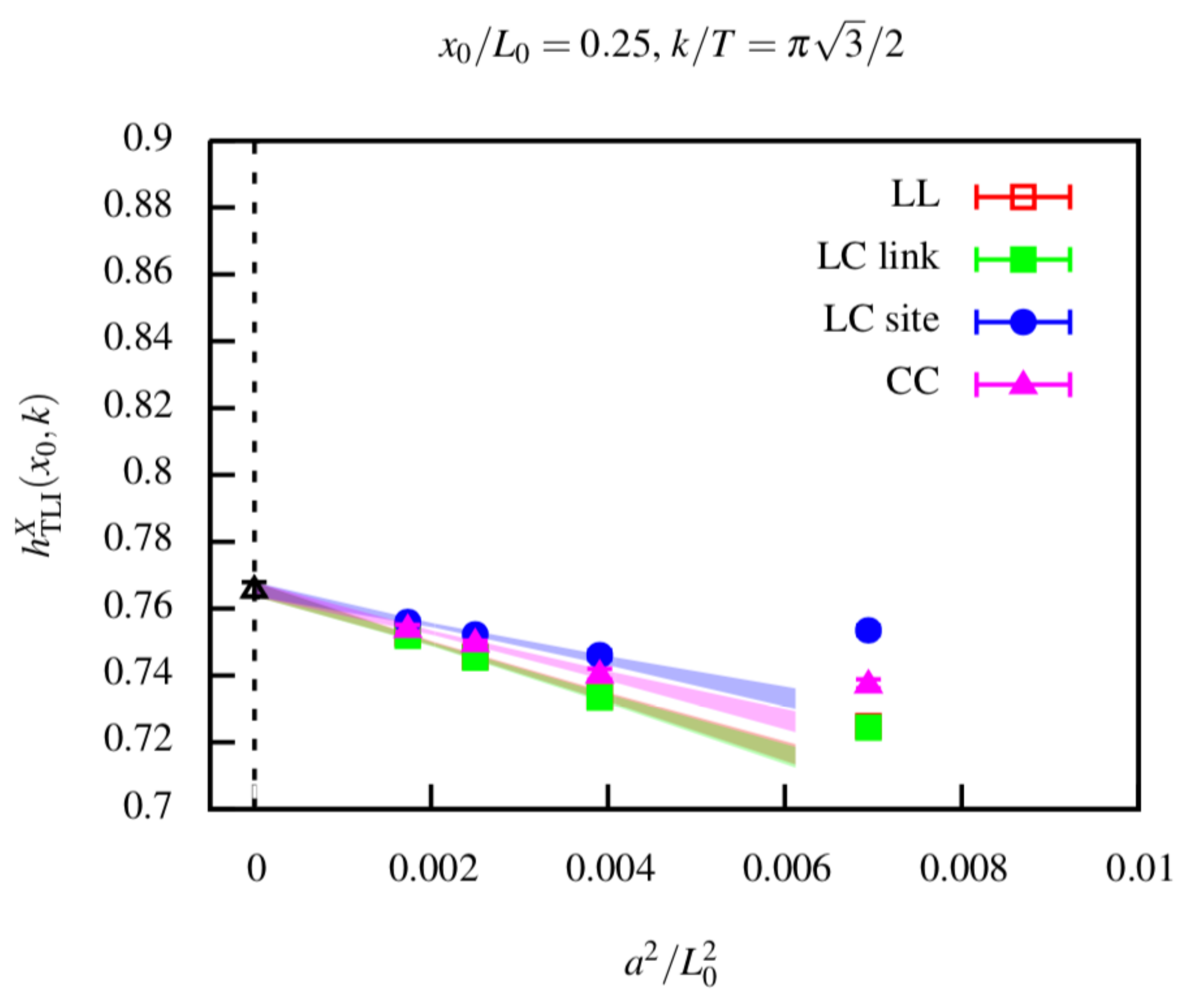}
\caption{Continuum extrapolation of $h^X(x_0, \abs{\textbf k}) \equiv \beta G^X(x_0, \abs{\textbf k}; \lambda = -2)/(2\chi_s)$, with $\chi_s \equiv \beta G_{00}(x_0,\textbf 0)$ the static susceptibility, for fixed momentum $\abs{\textbf k} \beta = \pi \sqrt 3/2$. $X = $ (LL, LC link, LC site, CC) represents the four different discretizations of the vector-current correlator defined in  \cite{Brandt:2017vgl}. Left: $h^{LL}(x_0, \abs{\textbf k})$ as a function of $x_0$, for three lattice ensembles, together with its continuum limit. In order to take the continuum limit at fixed Euclidean time, the correlators are interpolated with a piecewise cubic spline \cite{Brandt:2017vgl}. Right: simultaneous continuum extrapolation of the four independent discretizations, for $x_0/\beta = 0.25$. The coarsest ensemble is excluded from the continuum limit.}
\label{continuum_limit}
\end{figure}

We measure the correlator in equation~(\ref{corr}), but instead of the electromagnetic current we use the isovector vector current (which in view of our final results amounts to neglecting quark-disconnected contributions), and we use the linear combination of the Lorentz indices of equation~(\ref{rho_lambda}) with $\lambda = -2$. Using the local and exactly-conserved definitions of the vector current, we construct four independent discretizations of the vector-current correlator \cite{Brandt:2017vgl}, which are used for a simultaneous continuum extrapolation. Moreover, the correlators are tree-level improved by being multiplied with the ratio of the tree-level continuum correlator to the tree-level lattice one \cite{Brandt:2017vgl}. The correlators are measured at several Euclidean-time separations $x_0$ and projected to all spatial momenta consistent with $\abs{\textbf k} \leq 2 \pi T$, averaging over all possible orientations of the momentum $\textbf k$. To take the continuum limit at fixed Euclidean time, the correlators are interpolated as functions of $x_0$ with a piecewise cubic spline \cite{Brandt:2017vgl}. In our final analysis, we only use data with $x_0 \geq \beta/4$ and $\abs{\textbf k} < 2 \pi T$. An example of a continuum extrapolation is shown in figure \ref{continuum_limit}.

\section{Fits to Pad{\'e} ansatz}
\label{fits}

The determination of the spectral function from a finite set of noisy correlators is an ill-posed inverse problem, which we choose to regulate by proposing a model for the spectral function.
The properties of the $\lambda = -2$ spectral function motivate the choice of the Pad{\'e} ansatz \cite{Brandt:2017vgl}
\begin{equation}
\frac{\rho(\omega,\textbf k; \lambda = -2)}{\tanh(\omega \beta/2)} = \frac{A(1+B\omega^2)}{(\omega^2 + a^2)[(\omega + \omega_0)^2 + b^2][(\omega - \omega_0)^2 + b^2] } \; ,
\label{Pade}
\end{equation}
depending on two linear parameters $A$ and $B$, and three nonlinear ones $a$, $b$ and $\omega_0$, whose values are to be determined by fits to the lattice correlators.
To make our analysis more constraining, we want to fit simultaneously data with different momenta (a fixed-momentum analysis was conducted in \cite{Brandt:2017vgl}). With this aim, we express the non-linear parameters as functions of the momentum, using two different polynomial forms
\begin{equation}
a(k) = a_0 + a_2(k^n - k_0^n) \: , \quad b(k) = b_0 + b_2(k^n - k_0^n)  \: , \quad \omega_0(k) = \widetilde \omega_0 + \widetilde \omega_2(k^n - k_0^n) \: ,
\label{QL}
\end{equation}
a linear one with $n = 1$ and a quadratic one with $n = 2$. In the above equation, $k_0$ represents the smallest momentum included in the multi-momentum fit, and $k \equiv \abs{\textbf k}$. Having identified a set of $N_k$ momenta to include in the simultaneous fit, we perform a scan in the six-dimensional space of non-linear parameters $(a_0,a_2,b_0,b_2,\widetilde \omega_0,\widetilde \omega_2)$ and apply the following strategy:
\begin{enumerate}
\item Given a point $(a_0,a_2,b_0,b_2,\widetilde \omega_0,\widetilde \omega_2)$ in parameter space, we determine the linear coefficient $B$ (at fixed momentum, for all $N_k$ momenta) by imposing the sum rule~(\ref{sum_rule}).
\item For each momentum, and for all Euclidean-time separations $x_0$, we compute the correlator arising from the model spectral function (retaining $A$ as a free parameter) by using equation~(\ref{spectral_rep}). We point out that numerical integration is not the fastest way to compute the model correlator. Inserting the ansatz~(\ref{Pade}) into equation~(\ref{spectral_rep}), the integral can be solved analytically by contour integration in the complex-$\omega$ plane and finally be expressed as a linear combination of Lerch transcendents $\phi(e^{\pm i \> 2 \pi x_0},1,\frac{1}{2} + i \frac{\omega_p}{2 \pi})$, where $\omega_p$ are the poles of equation~(\ref{Pade}). With this method, we could perform scans over $O(10^8)$ points in parameter space at comparably low computational cost.
%
\item Given the lattice correlators in the continuum limit and the model correlators, we determine one value of $A$ for each of the $N_k$ momenta by $\chi^2$ minimization. We consider a fully-correlated $\chi^2$ with an $N_k N_{x_0} \times N_k N_{x_0}$ covariance matrix, where $N_{x_0} = 7$ is the number of Euclidean-time separations included in the analysis. In order to account for the residual cutoff effects, which may be larger than the smallest eigenvalues of the covariance matrix, we regularize the latter. Introducing two parameters $x$ and $y$ with values between 0 and 1, we define a regularized covariance matrix as follows: we identify $N_{x_0} \times N_{x_0}$ sub-blocks  $C^{(ij)}$ correlating data with fixed momenta $k_i$ and $k_j$ ($i,j = 1, \dots, N_k$), and obtain regularized sub-blocks $\tilde C^{(ij)}$ as: $\tilde C^{(ij)}_{x_0 x_0'} = (1-y) \delta^{ij} \hat C^{(ii)}_{x_0 x_0'} + y  \: \hat C^{(ij)}_{x_0 x_0'}$, where $ \hat C^{(ij)}_{x_0 x_0'} = (1-x) \delta_{x_0 x_0'} C^{(ij)}_{x_0 x_0} + x \: C^{(ij)}_{x_0 x_0'}$.
We will comment on the values of $x$ and $y$ used in our analysis in the next section.
\item We apply physically-motivated constraints to the parameters $a(k)$, $b(k)$ and $\omega_0(k)$. Given that relaxation times in the plasma cannot be arbitrarily long, we impose a lower bound to the imaginary part of the poles of the ansatz~(\ref{Pade}): $a(k), b(k) > \min (D_{\mathrm{strong}} \: k^2, D_{\mathrm{weak}}^{-1})$ \cite{Brandt:2017vgl}.
To set this bound, we have chosen two representative processes with large relaxation time: the diffusion of a charge-density perturbation in strongly-coupled $\pazocal N = 4$ SYM, with relaxation rate $D_{\mathrm{strong}} \: k^2 = (1/2 \pi T) k^2$, and the dissipation of a static current in weakly-coupled QCD, with leading-order relaxation rate $D_{\mathrm{weak}}^{-1} = 0.46 \: T$ \cite{Arnold:2003zc}.  We also impose the upper bounds: $a(k), b(k) < 2.5 \: k$, $\omega_0(k) < 3 \: k$,
motivated by the observation that, in NLO perturbation theory, the onset of the OPE-regime $\rho(\omega,k; \lambda = -2) \sim O(1/\omega^4)$  \cite{Brandt:2017vgl} is around $\omega \sim 20 \: T$ \cite{Jackson:2019yao}. Finally, to enforce spectral-function positivity for $\omega \leq k$, we require $A \geq 0$ and $B \geq -1/k^2$.
\end{enumerate}

\section{Results}

\begin{figure}[t]
\includegraphics[width=.5\textwidth]{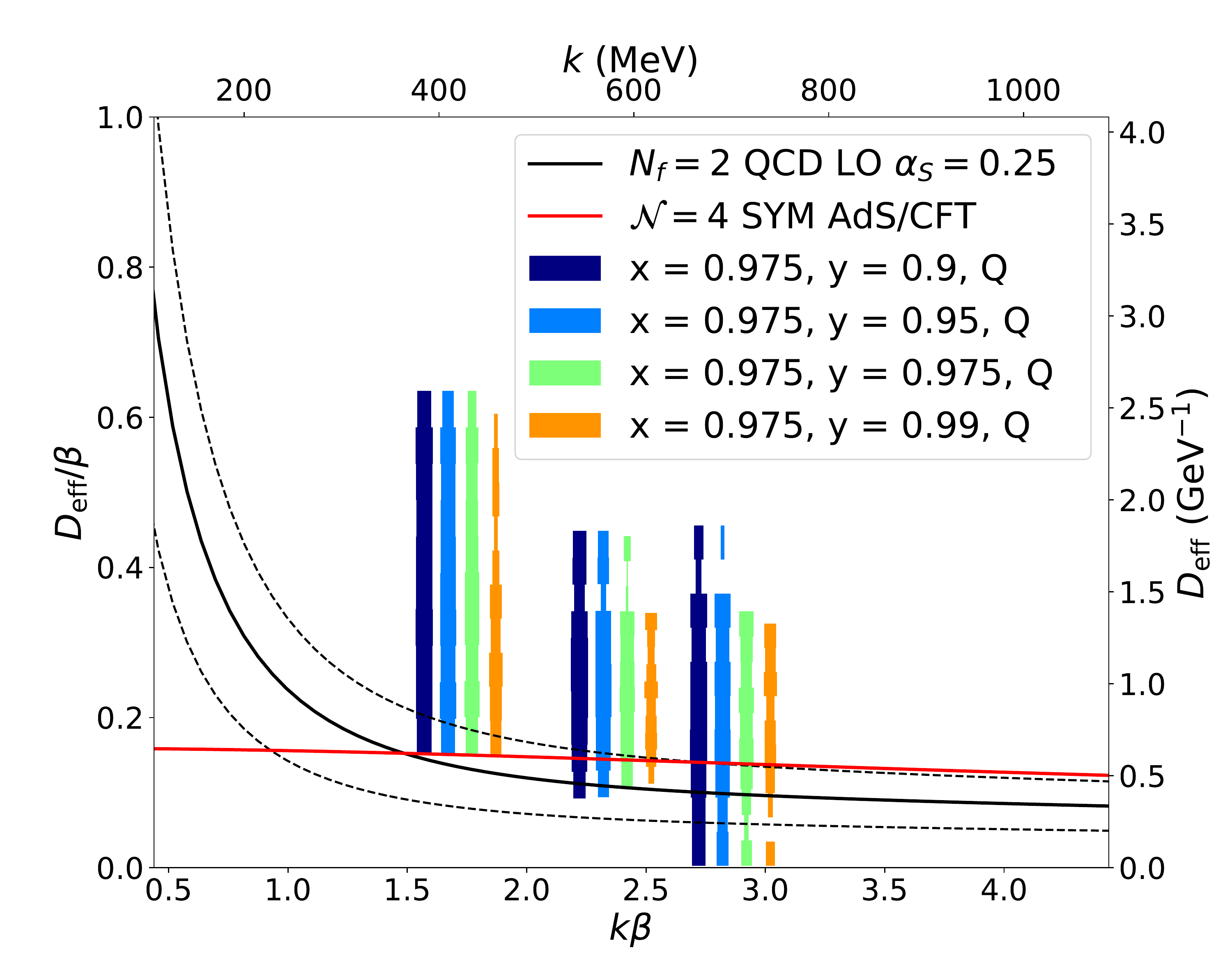}~\includegraphics[width=.5\textwidth]{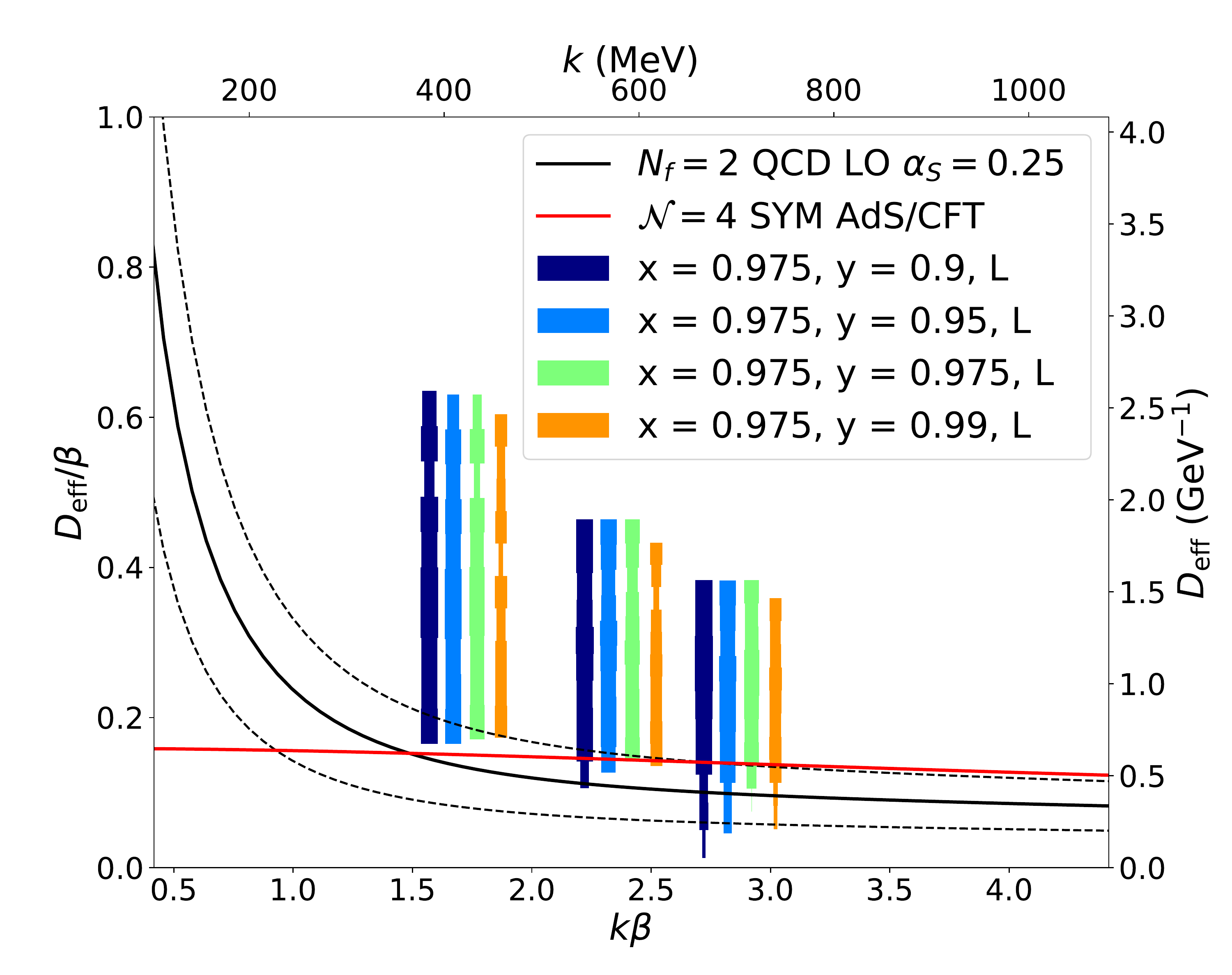}
\caption{A representative scan of parameter space for the momentum group $G1$, repeated for different values of $y$ and for $x = 0.975$. Acceptable $\Deff$ results are reported together with results from leading-order perturbation theory \cite{Arnold:2001ms} and from strongly-coupled $\pazocal N = 4$ SYM \cite{CaronHuot:2006te}. For each $\Deff$ bin, we consider the lowest $\chi^2$ found in it, and represent this information with lines of different thickness. Thicker lines correspond to lower $\chi^2$. As a consequence of this analysis, we choose the value $y=0.975$. For $y=1$, no acceptable $\chi^2$ was found with the linear ansatz (equation~(\ref{QL}) with $n=1$), and only one acceptable value with the quadratic ansatz (equation~(\ref{QL}) with $n=2$). Left: Quadratic ansatz. Right: Linear ansatz.}
\label{y}
\end{figure}

We divided the available momenta $k_n = n \: T \pi/2$ into three groups: $G1 = \{k_n \: \vert \: n^2 = 1,2,3\}$, $G2 = \{k_n \: \vert \: n^2 = 3,4,5,6,8\}$, $G3 = \{k_n \: \vert \: n^2 = 8,9,10,11,12,13,14\}$, and to each of the groups we applied the strategy described in section \ref{fits}. We present results for an observable which is proportional to the photon emission rate: the effective diffusion coefficient \cite{Ghiglieri:2016tvj}
\begin{equation}
\Deff \equiv \frac{\rho(\omega = k, k; \lambda = -2)}{4 \: k \: \chi_s} \; ,
\end{equation}
where $\chi_s \equiv \beta G_{00}(x_0,\textbf 0)$ is the static susceptibility. The result of our continuum extrapolation is $\chi_s/T^2 = 0.88(9)_{\mathrm{stat}}(8)_{\mathrm{syst}}$, to be compared with unity for non-interacting massless quarks.

As already observed in the fixed-momentum analysis \cite{Brandt:2017vgl}, the $\chi^2$ has no clear global minimum. In the case of multi-momentum fits, we observe many almost degenerate local minima in the $\chi^2$-landscape, which motivates us to retain all solutions with acceptable $\chi^2$ as possible solutions compatible with the lattice data. We choose a minimum $p$-value $p_0 = 0.32$, and we consider as acceptable all solutions with $p \geq p_0$. 
As discussed in section \ref{fits}, the covariance matrix has been regularized. In order to fix the regularization parameters $x$ and $y$, we chose a representative scan of parameter space and we repeated it for different values of $x$ and $y$. We fixed at first $y=0$ and repeated the scan for several values of $x$. We chose for our analysis the largest $x$ such that $\Deff$ results are stable under small changes in $x$. For the group G1 this value is $x = 0.975$. Having fixed $x$, we repeated the same procedure to choose an appropriate value of $y$. For G1, this corresponds to $y=0.975$, as shown in figure \ref{y}. 

\begin{figure}[t]
\center
\includegraphics[width=.505\textwidth]{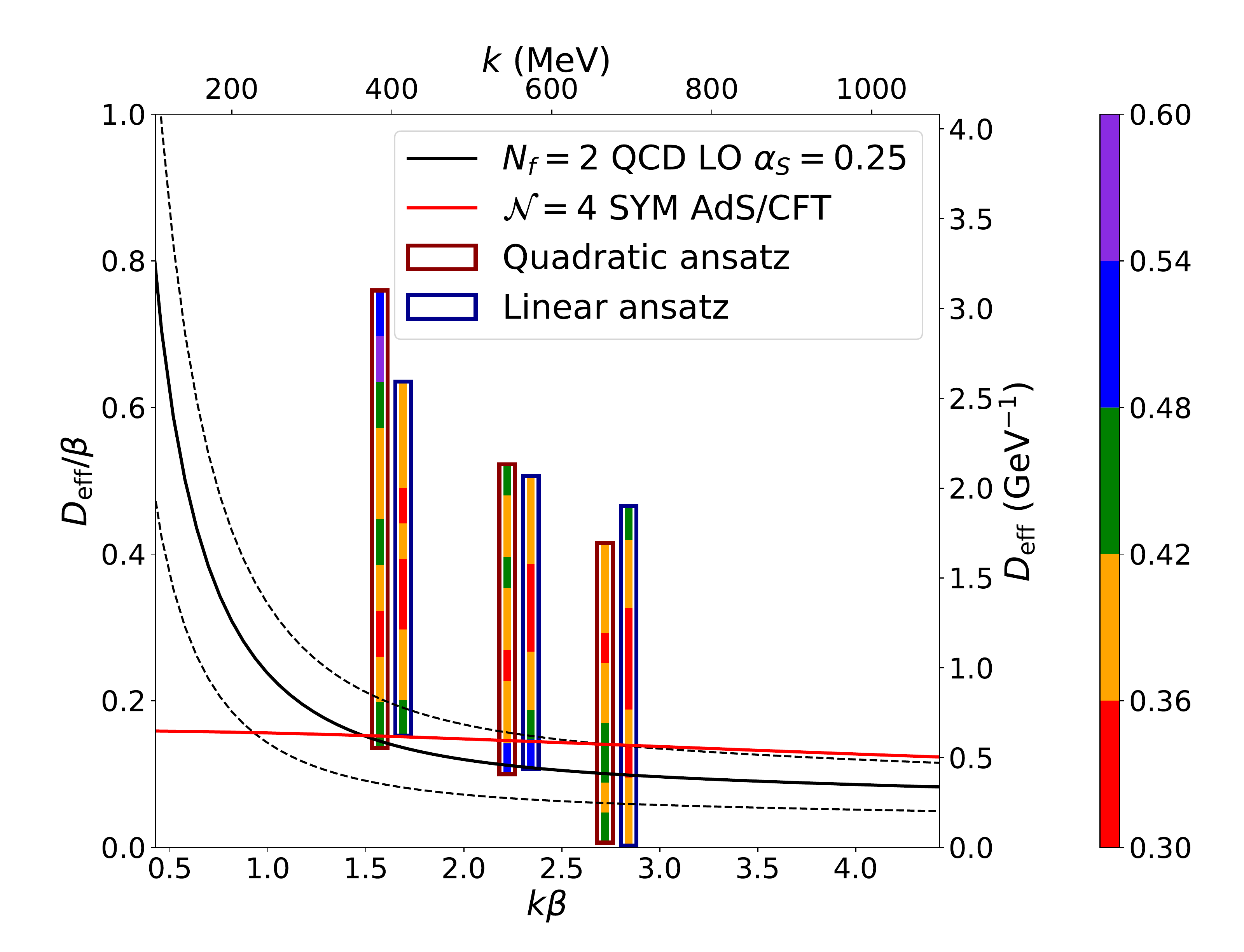}~\includegraphics[width=.495\textwidth]{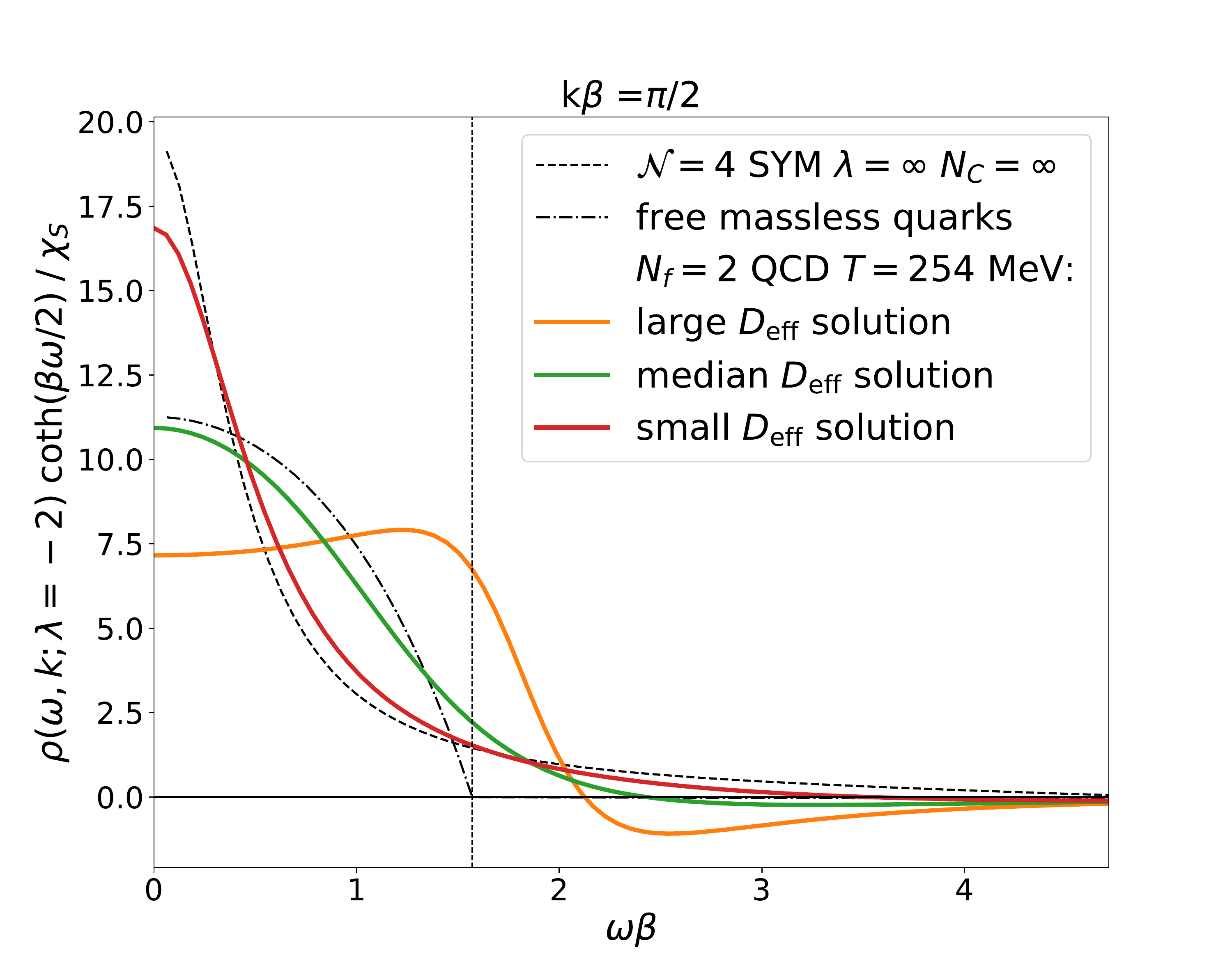}
\caption{Left: Acceptable $\Deff$ results for the group $G1$ from our finest scans of parameter space, together with results from leading-order perturbation theory \cite{Arnold:2001ms} and from strongly-coupled $\pazocal N = 4$ SYM \cite{CaronHuot:2006te}. We represent our results with a histogram, where each $\Deff$ bin is colored according to the minimum $\chi^2$ per degree of freedom found in it. The quadratic ansatz (equation~(\ref{QL}) with $n=2$) and the linear ansatz (equation~(\ref{QL}) with $n=1$) give rise to comparable results. Right: Illustrative examples of spectral functions with $k = \pi/2 \: T$, corresponding to low, median and large values of $\Deff$, together with the spectral functions of non-interacting quarks and of strongly-coupled $\pazocal N = 4$ SYM. The vertical line corresponds to $\omega = k$.}
\label{final}
\end{figure}

\begin{figure}[t]
\vspace{-0.6cm}
\center
\includegraphics[width=.9\textwidth]{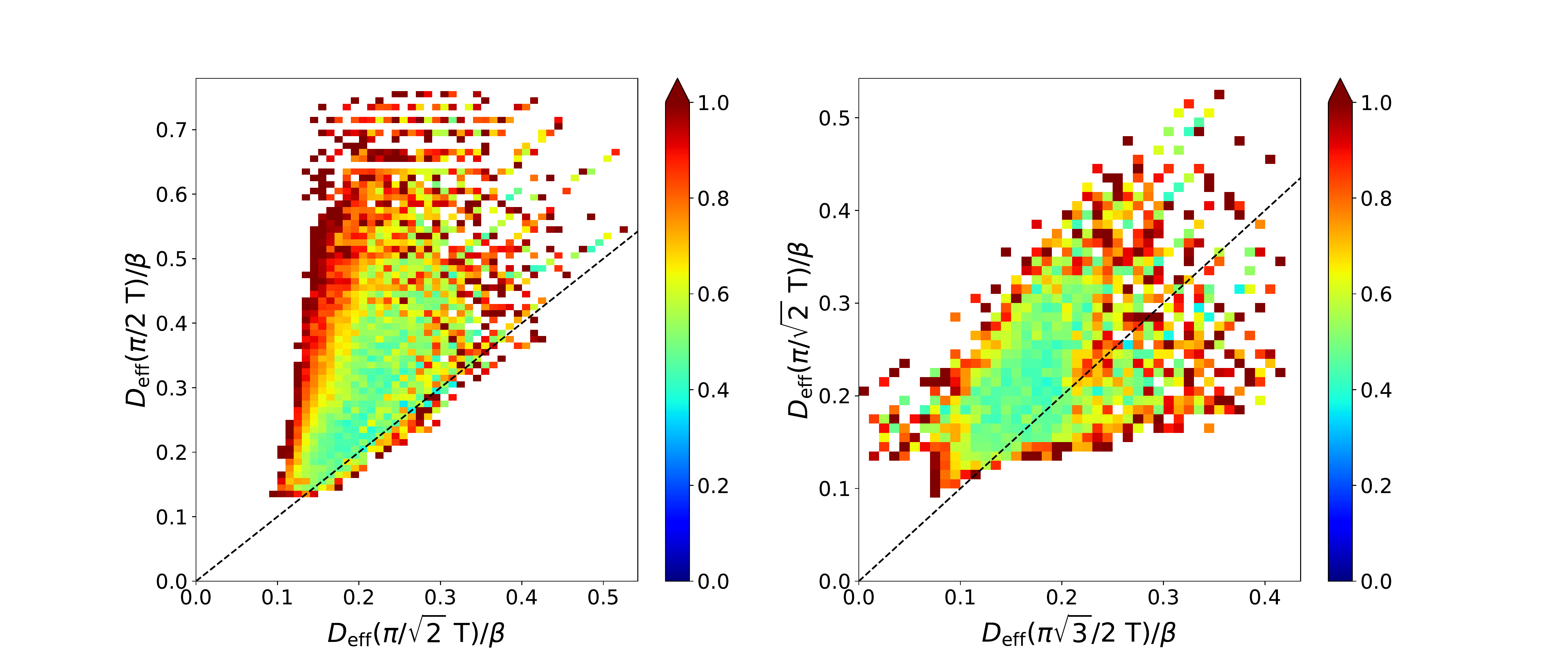}
\caption{$\Deff(\pi \: T/2)$ vs. $\Deff(\pi \sqrt 2 \: T/2)$ (left) and $\Deff(\pi\sqrt 2 \: T/2)$ vs.  $\Deff(\pi \sqrt 3 \: T/2)$ (right). Acceptable $\Deff$ values are represented with a histogram, where each bin is colored according to the minimum $\chi^2$ per degree of freedom found in it. The diagonal is represented as a dashed line.}
\label{ratio}
\end{figure}

The results of our finest scans of parameter space for momenta in $G1$ are shown in the left panel of figure \ref{final}, while the right panel shows examples of spectral functions.
As an advantage of simultaneously analyzing multiple momenta, our results carry information on the momentum dependence of $\Deff$. A possible way to visualize this information is shown in figure~\ref{ratio}. We observe a clear correlation between values of $\Deff$ at subsequent momenta.

In the future, we will present results for the groups $G2$ and $G3$. Moreover, we plan on studying the $\lambda = -2$ correlator at fixed virtuality, \textit{i.e.}~projected to imaginary spatial momentum $k = i \omega$, which exclusively probes the photon rate, rather than receiving contributions from the whole $(\omega,k)$ dependence of the spectral function \cite{Meyer:2018xpt}.


\vspace{0.5cm}
\noindent \textbf{Acknowledgments:} Computing resources for this project were provided by the Clover and Himster~II HPC clusters at Helmholtz-Institut Mainz and Mogon II at JGU Mainz. We acknowledge the use of computing time on the JUGENE and JUQUEEN computers at FZ J{\"u}lich allocated under the grant HMZ21. This work was supported in part by DFG Grant ME 3622/2-2, by the European Research Council (ERC) through Grant Agreement No. 771971-SIMDAMA and by DFG - project number 315477589 - TRR 211.

\vspace{-0.2cm}
\bibliographystyle{JHEP}
\bibliography{proceedings_Arianna_Toniato}

\providecommand{\href}[2]{#2}\begingroup\raggedright\begin{thebibliography}{1}

\bibitem{Brandt:2017vgl}
B.~B. Brandt, A.~Francis, T.~Harris, H.~B. Meyer and A.~Steinberg, \emph{{An
  estimate for the thermal photon rate from lattice QCD}},
  \href{https://doi.org/10.1051/epjconf/201817507044}{\emph{EPJ Web Conf.}
  {\bfseries 175} (2018) 07044}
  [\href{https://arxiv.org/abs/1710.07050}{{\ttfamily 1710.07050}}].

\bibitem{McLerran:1984ay}
L.~D. McLerran and T.~Toimela, \emph{{Photon and Dilepton Emission from the
  Quark - Gluon Plasma: Some General Considerations}},
  \href{https://doi.org/10.1103/PhysRevD.31.545}{\emph{Phys. Rev.} {\bfseries
  D31} (1985) 545}.

\bibitem{Fritzsch:2012wq}
P.~Fritzsch, F.~Knechtli, B.~Leder, M.~Marinkovic, S.~Schaefer, R.~Sommer
  et~al., \emph{{The strange quark mass and Lambda parameter of two flavor
  QCD}}, \href{https://doi.org/10.1016/j.nuclphysb.2012.07.026}{\emph{Nucl.
  Phys.} {\bfseries B865} (2012) 397}
  [\href{https://arxiv.org/abs/1205.5380}{{\ttfamily 1205.5380}}].

\bibitem{Arnold:2003zc}
P.~B. Arnold, G.~D. Moore and L.~G. Yaffe, \emph{{Transport coefficients in
  high temperature gauge theories. 2. Beyond leading log}},
  \href{https://doi.org/10.1088/1126-6708/2003/05/051}{\emph{JHEP} {\bfseries
  05} (2003) 051} [\href{https://arxiv.org/abs/hep-ph/0302165}{{\ttfamily
  hep-ph/0302165}}].

\bibitem{Jackson:2019yao}
G.~Jackson and M.~Laine, \emph{{Testing thermal photon and dilepton rates}},
  \href{https://arxiv.org/abs/1910.09567}{{\ttfamily 1910.09567}}.

\bibitem{Arnold:2001ms}
P.~B. Arnold, G.~D. Moore and L.~G. Yaffe, \emph{{Photon emission from quark
  gluon plasma: Complete leading order results}},
  \href{https://doi.org/10.1088/1126-6708/2001/12/009}{\emph{JHEP} {\bfseries
  12} (2001) 009} [\href{https://arxiv.org/abs/hep-ph/0111107}{{\ttfamily
  hep-ph/0111107}}].

\bibitem{CaronHuot:2006te}
S.~Caron-Huot, P.~Kovtun, G.~D. Moore, A.~Starinets and L.~G. Yaffe,
  \emph{{Photon and dilepton production in supersymmetric Yang-Mills plasma}},
  \href{https://doi.org/10.1088/1126-6708/2006/12/015}{\emph{JHEP} {\bfseries
  12} (2006) 015} [\href{https://arxiv.org/abs/hep-th/0607237}{{\ttfamily
  hep-th/0607237}}].

\bibitem{Ghiglieri:2016tvj}
J.~Ghiglieri, O.~Kaczmarek, M.~Laine and F.~Meyer, \emph{{Lattice constraints
  on the thermal photon rate}},
  \href{https://doi.org/10.1103/PhysRevD.94.016005}{\emph{Phys. Rev.}
  {\bfseries D94} (2016) 016005}
  [\href{https://arxiv.org/abs/1604.07544}{{\ttfamily 1604.07544}}].

\bibitem{Meyer:2018xpt}
H.~B. Meyer, \emph{{Euclidean correlators at imaginary spatial momentum and
  their relation to the thermal photon emission rate}},
  \href{https://doi.org/10.1140/epja/i2018-12633-0}{\emph{Eur. Phys. J.}
  {\bfseries A54} (2018) 192}
  [\href{https://arxiv.org/abs/1807.00781}{{\ttfamily 1807.00781}}].

\end{thebibliography}\endgroup

\end{document}